# Incremental Deployment of Network Monitors Based on Group Betweenness Centrality

Shlomi Dolev, Yuval Elovici, Rami Puzis, Polina Zilberman

**Abstract**

In many applications we are required to increase the deployment of a distributed monitoring system on an evolving network. In this paper we present a new method for finding candidate locations for additional deployment in the network. This method is based on the Group Betweenness Centrality (GBC) measure that is used to estimate the influence of a group of nodes over the information flow in the network. The new method assists in finding the location of *k* additional monitors in the evolving network, such that the portion of additional traffic covered is at least $\left(1 - \frac{1}{e}\right)$ of the optimal.

## 1. Introduction

Distributed systems designed to work on wide area networks are commonly used for network monitoring tasks such as traffic measurements and intrusion detection and prevention. Networks tend to evolve over time and there is a constant need to update the deployment of monitors in order to maintain the system effectiveness. In many cases relocating the deployment is not practical both from economical and from technical perspectives (for example when deployment involves infrastructure adjustments). In these cases, there is a need to find candidate locations for additional deployment of network monitors.

Several methods were proposed for optimizing the location of network monitors by maximizing the number of network flows covered. Suh et al. [1] proposed approximate solutions for several monitor location problems using greedy heuristics. Chaudet at el. [2] presented a framework, based on Mixed Integer Programming, for solving monitor placement problems with additional "must deploy" constraint. The methods presented in [1, 2] operate on the collection of paths utilized by each source-target flow. Therefore, they are most applicable when there are few arbitrary paths for each source-target flow. These methods are evaluated on commercial internet service provider networks and Point-of-Presence topologies.

The algorithm presented in this paper reduces the computational effort required for optimizing placement of passive traffic monitors in cases where network flows utilize all shortest paths from source to target. We avoid enumerating shortest paths by exploiting efficient computation of Shortest Path Group Betweenness Centrality (GBC) [3–6], which is roughly defined as the total fraction of shortest paths that traverse at least one member of the group. Betweenness centrality can easily be adapted to consider variable communication patterns in the form of traffic matrices [7]. Puzis et al. describe in [8] a greedy algorithm that, at every stage, chooses the candidate that contributes the most to the GBC of the already chosen candidates. The algorithm presented here, is a generalization of this scheme which also supports "must deploy" and "can not deploy" constraints. It allows optimizing monitors' deployment on evolving networks whose topology is changing over time.

The following example describes a potential application of the algorithm. Let $G(V,E)$ be a communication network and $D \subseteq V$ be a set of nodes chosen for deployment of network monitors. Assume that the network has evolved since the initial deployment into $G'(V',E')$ and $D \subseteq V'$. Assume also that there is a budget for additional $k$ monitors chosen from a set of candidate locations $C \subseteq V'$. The additional $k$ monitors should be deployed to cover the largest possible portion of a traffic that is not covered already by $D$. The proposed algorithm receives $G'$, $D$, $C$, and $k$ as an input. It returns a group of nodes $M \subseteq V'$ such that $D \subseteq M$ and the portion of additional traffic covered by $M$ is at least $\left(1 - \frac{1}{e}\right)$ of the optimal. Note that this approximation factor holds only when network flows utilize all shortest paths from source to target with equal probability. Nevertheless, it has been shown in the literature that betweenness predicts well the traffic load on nodes in communication networks with conventional shortest path routing [9].

The rest of the paper is structured as follows. In section 2 we present the new computational method. In section 3 we prove that the portion of additional traffic covered is at least $\left(1 - \frac{1}{e}\right)$ of the optimal. Section 4 presents experimental results and section 5 concludes the paper with a summary.

## 2. Incremental Deployment Method

### *2.1 Definitions of Shortest Path Betweenness Centrality Measures*

Betweenness Centrality (BC) accounts for shortest paths between all pairs of nodes in a network. Let *s* and *t* be two nodes in a network. $\sigma_{s,t}$ denotes the total number of different shortest paths that connect *s* and *t*. Let *v* be a node that resides on a shortest path between *s* and *t*. $\sigma_{s,t}(v)$ denotes the number of shortest paths from *s* to *t* that pass through *v*.

$$\sigma_{s,t}(v) = \begin{cases} \sigma_{s,v} \cdot \sigma_{v,t} & d(s,t) = d(s,v) + d(v,t) \\ 0 & otherwise \end{cases} \quad (1)$$

where *d*(*x, y*) is the distance between nodes *x* and *y*. For example, in Figure 1 $\sigma_{1,4} = 3$ since there are three shortest paths from 1 to 4. For computing $\sigma$ and distance matrices for a network *G* with *n* nodes and *m* edges, we use the algorithm presented by Brandes [10]. The computational complexity of this algorithm is *O*(*nm*) and does not depend on the actual number of shortest paths between each pair of nodes. We exploit this feature of the algorithm to devise a strategy for deployment of network monitors in networks where traffic can flow through any shortest path from source to target.

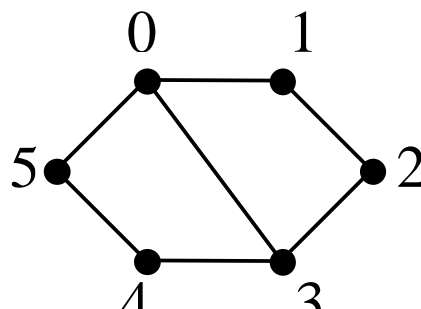

Figure 1: Sample network with a network flow between any two nodes. Nodes 0 and 3 have the highest traffic load.

BC of node $v \in V$ represents the total influence that *v* has on communications between all possible pairs of nodes in the network. BC of node *v* is

$$BC(v) = \sum_{s,t \in V | s \neq t} \left( \frac{\sigma_{s,t}(v)}{\sigma_{s,t}} \right) \quad (2)$$

where the fraction $\frac{\sigma_{s,t}(v)}{\sigma_{s,t}}$ represents the influence of *v* on the communication between *s* and *t,* and where $\sigma_{x,x} = 1$ which results in $\sigma_{s,t}(t) = \sigma_{s,t}(s) = \sigma_{s,t}$. As an example, let us recall Figure 1: $BC(2) = 2 \times \left(1 + \frac{1}{2} + 1 + 1 + 1 + 1 + \frac{1}{3}\right) = 11.667$

GBC is a natural extension of BC of individual nodes [4]. Let $S \subseteq V$ be a group of nodes. *GBC*(*S*) stands for the total fraction of shortest paths between all pairs of nodes that pass through at least one member of the group *S*. Let $\ddot{\sigma}_{s,t}(S)$ be the number of shortest paths between *s* and *t* that traverse at least one member of the group *S*. GBC of group *S* is

$$GBC(S) = \sum_{s,t \in V | s \neq t} \left( \frac{\ddot{\sigma}_{s,t}(S)}{\sigma_{s,t}} \right) \qquad (3)$$

For instance, in Figure 1 it holds that $GBC(\{2, 3\}) = 20.3333$.

Path Betweenness (PB) centrality [3], generalizes the concept of single node Betweenness to Betweenness of sequences of nodes $S = (v1, v2, .., vk)$. $PB(S)$ stands for the total fraction of shortest paths between all pairs of nodes that traverse all nodes in $S$. Let $\tilde{\sigma}_{s,t}(S)$ be the number of shortest paths between $s$ and $t$ that traverse all members of $S$. PB of $S$ is:

$$PB(S) = \sum_{s,t \in V | s \neq t} \left( \frac{\tilde{\sigma}_{s,t}(S)}{\sigma_{s,t}} \right) \qquad (4)$$

Again, the network in Figure 1 will serve as an example: $PB(\{2,3\}) = 2 \times \left(\frac{1}{2} + \frac{1}{3} + 1 + 1 + \frac{2}{3}\right) = 7$.

Note that $BC(v) = GBC(\{v\}) = PB(v,v)$ by definition.

***2.2 Two-phase Algorithm for Incremental Deployment***

Let $G = (V, E)$, with $|V| = n$ nodes and $|E| = m$ edges, be a communication network, where a route from source $s$ to target $t$ is some shortest path chosen uniformly out of all shortest paths from $s$ to $t$. Let $D$ be a group of nodes that has some control over the traffic in $G$. Let $C \subseteq V$ be a set of candidate locations for additional deployment. We assume without the loss of generality that $C \cap D = \phi$. We want to find a group of $|D|+k$ nodes denoted by $M$ where $M \subseteq C \cup D$, $D \subseteq M$, and $M$ has maximal control over the traffic in $G$.

We define an algorithm that in the first phase calculates $GBC_G(D)$ and in the second phase it constructively finds the additional $k$ nodes. In every step during the second phase, the algorithm chooses the next best candidate $v$ according to its contribution to GBC of the current group. Initially $M = \phi$, after the first phase $M=D$, and after the second phase $|M|=|D|+k$. Each time we add a node to $M$ we update the following data structure:

- $\sigma^M$ – $l \times l$ matrix, $|C \cup D| = l$, whose elements $\sigma^M_{s,t} : (s,t \in C \cup D)$ store the number of shortest paths between $s$ and $t$ that do not traverse any node in $M$. The initial values of $\sigma^M_{s,t}$ (for $M = \phi$) are equal to $\sigma_{s,t}$.
- $PB^M$ – $l \times l$ matrix, whose elements $PB^M(x, y) : (x, y \in C \cup D)$ store the PB of pair $(x, y)$ disregarding the shortest paths that traverse at least one node in $M$. Initial values of $PB^M(x, y)$ (for $M = \phi$) are equal to $PB(x, y)$. The values $PB(x, y)$ are computed only for $x, y \in C \cup D$.

*GBC(M)* is the total fraction of the shortest paths that traverse at least one node in *M*. $PB^M(v,v)$ is the total fraction of shortest paths that traverse *v* excluding shortest paths that traverse at least one node in *M*. Therefore, $GBC(M) + PB^M(v,v) = GBC(M \cup \{v\})$.

Algorithm 1 presents a procedure for updating $\sigma^M$ and $PB^M$ matrices to exclude paths that traverse *v*. Algorithm 1 loops through all nodes in $C \cup D$. In Line 2 $\sigma^M_{x,y}$ is decreased by $\sigma^M_{x,y}(v)$, removing paths that traverse *v*. In the special case $x = y \neq v$, Line 3 subtracts from $PB^M(x,x)$ (the current contribution of *x*) paths that also traverse *v* in both directions *x* to *v* and *v* to *x*. In a general case, Line 4 uses $\sigma^M$ to calculate the value of $PB^{M\cup\{v\}}(x, y)$ by removing from $PB^M(x, y)$ paths that include *v*. Note that there is a path from *x* to *y* that traverses *v* if and only if *d*(*x*, *v*) + *d*(*v*, *y*) = *d*(*x*, *y*). If there is a shortest path traversing x, y, and v (first *x* and then *y*), one of three options is considered: *v* is before *x*, *v* is after *y*, or *v* is between *x* and *y*. In general, a node *u* is before node *w* on the path from source *s* to destination *t*, if the distance of *u* from *s* is less than the distance of *w* from *s*). If no such shortest path exists, $\sigma^M_{w_1,w_3}(w_2) = 0$ for any assignment of $w_1, w_2$ *and* $w_3$.

*Input:* $C, D, \sigma^M, PB^M, v$

*Output:* $\sigma^{M\cup\{v\}}, PB^{M\cup\{v\}}$

1: *for each* $x, y \in C \cup D$ *repeat* 2 – 4

2: $\sigma^{M\cup\{v\}}_{x,y} = \sigma^M_{x,y} - \sigma^M_{x,y}(v)$

3: *if* $x = y \neq v$

// *update* $PB^M$ *considering* $(v,x)$ *and* $(x,v)$

$PB^{M\cup\{v\}}(x,x) =$

$= PB^M(x,x) - PB^M(v,x) - PB^M(x,v)$

4: *else if there is shortest path containing x,y, and v assign* $x, y, v$ *to* $w_1, w_2$, *and* $w_3$ *where* $w_2$ *is between* $w_1$ *and* $w_3$.

$PB^{M\cup\{v\}}(x,y) =$

$$= PB^M(x,y) - \frac{\sigma^M_{w_1,w_3}(w_2)}{\sigma^M_{w_1,w_3}} PB^M(w_1,w_3)$$

Algorithm 1: Update procedure.

*Input:* $D, C, \sigma, PB^M, k : (|C \cup D| = l)$

*Output: a group of k nodes*

// *Initial phase*

1: $M \longleftarrow \phi$

2: $\forall x, y \in C \cup D, \sigma^M{}_{x,y} \longleftarrow \sigma_{x,y}$

3: $\forall x, y \in C \cup D, PB^M(x, y) \longleftarrow PB(x, y)$

// *First phase*

4: *for each* $v \in D$

5: *update*$(C, D, M, \sigma^M, PB^M, v)$

6: $M \longleftarrow M \cup \{v\}$

// *Second phase*

7: *repeat k times*:

8: *find* $v \in C$ *with maximal* $PB^M(v,v)$

9: *update*$(C, D, M, \sigma^M, PB^M, v)$

10: $M \longleftarrow M \cup \{v\}$

*return* $M \setminus D$

Algorithm 2: Two-Phase incremental deployment

Algorithm 2 receives as an input the existing deployment *D*, the set of candidate locations *C*, $\sigma^M$ and $PB^M$ matrices and the number of additional monitors to be deployed *k*. Lines 1-4 initialize its data structure. The first phase (Lines 4-6) goes through all nodes in *D* preparing the $\sigma^M$ and $PB^M$ matrices for the second phase. In the second phase we iteratively add to *M k* nodes, choosing each time the node with highest contribution to the *GBC* of *M*. The contribution of a node *v* to *GBC*(*M*) is $PB^M(v,v)$. The contribution of the first node is therefore, $PB^{\phi}(v_1,v_1)$. The second node contributes $PB^{\{v_1\}}(v_2,v_2)$, and so on. We continue updating the matrices $\sigma^M$ and $PB^M$ with respect to the growing *M* by removing shortest paths that traverse *v*. Assume that the existing deployment in the network depicted in Figure 1 is *D* = {1}. According to Algorithm 2 the first node to be added to the deployment is 3.

Let $|C \cup D| = l$. The running time of Algorithm 2 scales as $|M|l^2$ ($|M|$ calls to the "update" procedure which updates $l^2$ entries each one in $O(1)$). The algorithm requires $l^2$ entries of the $\sigma$ and *PB* matrices. The complexity of these matrices' computation is $max\{O(nl^2), O(nm)\}$ [3]. Since $O(nl^2)$ dominates $O(|M|l^2)$ the total time spent on constructing *M* is $O(nm)$ when $l \le \sqrt{m}$ or $O(nl^2)$ when $\sqrt{m} \le l$.

## 3. The Approximation Factor

In this section we prove that Algorithm 2 is $\left(1 - \frac{1}{e}\right)$ approximation for the problem of finding a set of *k* nodes with maximal contribution to the *GBC* of a given set of nodes *D*. The proof is inspired by the well known greedy approximation algorithm for max-k variant of the set-cover problem [11]. The following proposition was proved in [3]:

PROPOSITION 1: *After each execution of the "update" procedure,* $PB^M(x,y)$ *is the Path Betweenness of (x, y) excluding shortest paths that traverse at least one node in M.*

In particular the above proposition implies that $BC^M(v) = PB^M(v,v)$ is the contribution of *v* to the *GBC* of the group of nodes *M*.

Through the end of this section we focus on the second phase of Algorithm 2. Let $M_i = D \cup \{v_1, \ldots, v_i\}$ be the set of nodes located by the algorithm during $i \le k$ iterations of the second phase and $v_i$ be the node chosen by the algorithm in Line 8. In particular, $M_0 = D$ and $v_1$ is the node with the highest contribution to the GBC of *D*.

PROPOSITION 2: *Let* $M' \subseteq C$ *be a set of k nodes with maximal contribution to the GBC of D, it holds that:* $k \cdot BC^{M_{i-1}}(v_i) \geq GBC^{M_0}(M') - GBC^{M_0}(M_{i-1})$ *where* $GBC^{M_0}(X)$ *denotes the contribution of X to the GBC of D.*

PROOF: $v_i$ is the node with the maximal contribution to the GBC of $M_{i-1}$ in particular $BC^{M_{i-1}}(v_i) \geq BC^{M_{i-1}}(v')$ for each $v'$ in $M'$. It was proved in [8] that the sum of contributions of any set of nodes to the GBC of $M_{i-1}$ is greater than or equal to their joint contribution. Therefore, it holds that: $k \cdot BC^{M_{i-1}}(v_i) \geq \sum_{v' \in M'} BC^{M_{i-1}}(v') \geq GBC^{M_{i-1}}(M')$.

$GBC^{M_{i-1}}(M')$ accounts for all shortest paths that pass through $M'$ but do not pass through $M_{i-1}$. Some of the shortest paths that pass through $M_{i-1} \setminus M_0$ are accounted for by $GBC^{M_0}(M')$ but not all. For instance, a path that passes through two vertices $u \in M_{i-1} \setminus M_0$ and $w \in M_{i-1} \cap M_0$ is not accounted for by $GBC^{M_0}(M')$. Therefore, we end up with the following inequality: $k \cdot BC^{M_{i-1}}(v_i) \geq GBC^{M_{i-1}}(M') \geq GBC^{M_0}(M') - GBC^{M_0}(M_{i-1})$.

□

PROPOSITION 3: *Algorithm 2 returns a set of nodes whose contribution to the GBC of D is at least* $\left(1 - 1/e\right) \approx 0.632$ *of the optimum.*

PROOF: Note that the contribution of $M_i$ to the GBC of *D* is trivially equal to the contribution of $M_i \setminus D$ to the GBC of *D*. $BC^{M_{i-1}}(v_i)$ is the marginal contribution of $v_i$ to the GBC of $M_{i-1}$ (where $M_0$ is ground zero). Therefore, summing the respective marginal contributions of the *i-1* nodes found in Line 8 ($BC^{M_{i-1}}(v_i)$) results in $GBC^{M_0}(M_{i-1})$. Let $M' \subseteq C$ be the set of nodes with maximal contribution to the GBC of *D*. According to Proposition 2 it holds that:

$$BC^{M_{i-1}}(v_i) \geq \frac{GBC^{M_0}(M') - \sum_{j=1}^{i-1} BC^{M_{j-1}}(v_j)}{k}.$$

Therefore, (similarly to Proposition 5.1 in [11])

$$\begin{aligned} GBC^{M_0}(M) &= \sum_{i=1}^{k} BC^{M_{i-1}}(v_i) \\ &\geq GBC^{M_0}(M') \cdot (1 - (1 - 1/k)^k). \\ &\geq GBC^{M_0}(M') \cdot (1 - 1/e) \end{aligned}$$

□

## 4. Deployment of monitoring devices

In this section we present a comparison of two strategies for updating the deployment of a monitoring system over an evolving network while preserving the system effectiveness. In the first strategy, a new deployment will be found in every time period regardless of the previous deployments. In the second, a search for additional deployment will be performed assuming that devices that were already deployed cannot be relocated.

We have performed experiments on evolving networks created using the BA model [12], with various average degrees. We assumed that monitoring devices should intercept 95% of the flows. The difference between the number of devices deployed according to the two strategies averaged over 20 networks is presented in Fig. 2. Each network evolved from 100 nodes to 2000 and the monitors' deployment was recalculated after each addition of 100 nodes. We can see that the absolute penalty (expressed in the number of additional monitors) for not relocating monitors that were already deployed increases as network grows. However, the relative size of such additional deployment is 2% on average and at most 6% in all simulations.

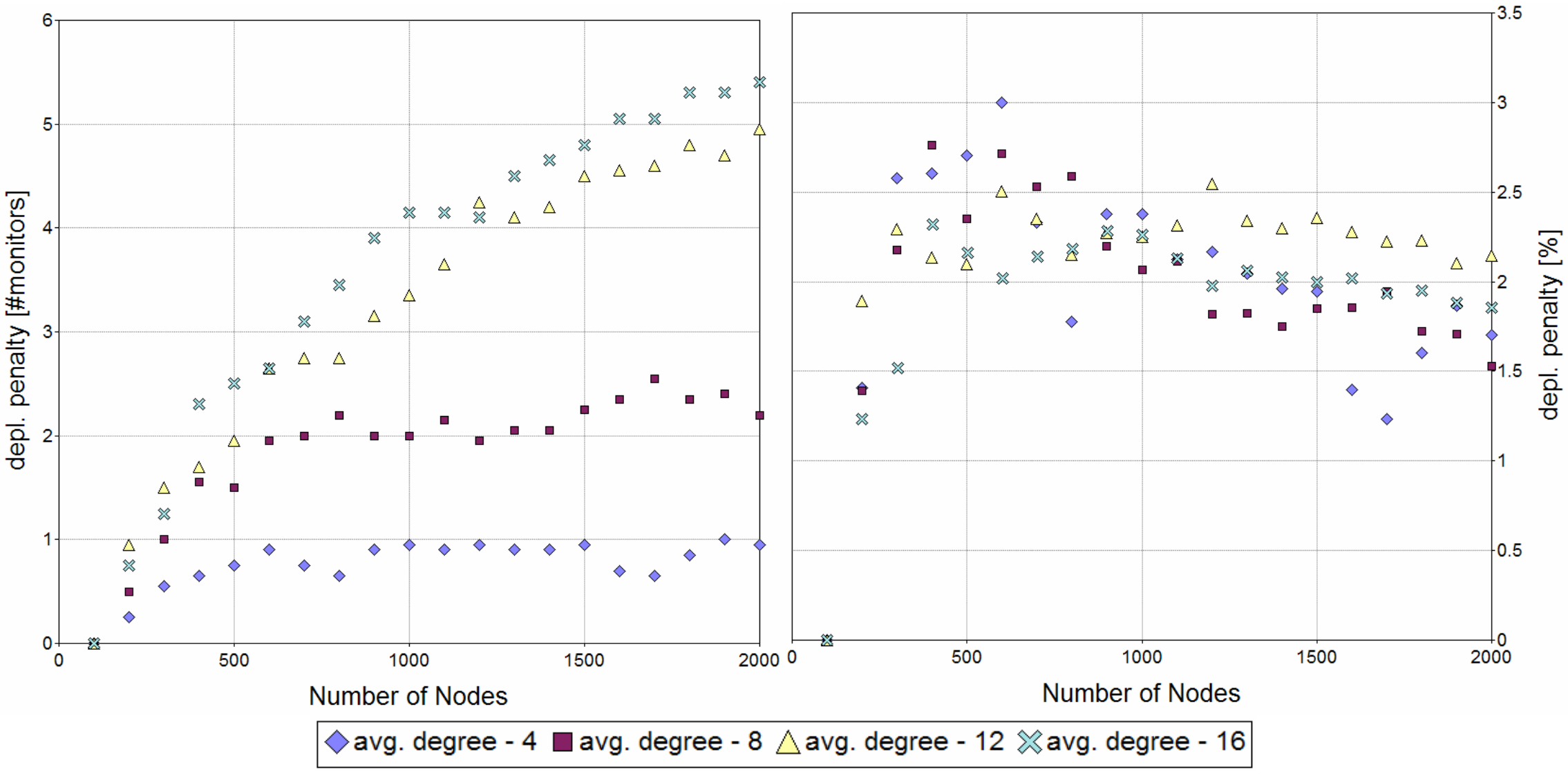

**Figure** 2: The penalty for not relocating network monitors in terms of the absolute (left) and relative (right) number of monitors for evolving BA networks with different average degrees.

## 5. Summary

In this paper we presented a new method for finding candidate locations for additional deployment of network monitors in evolving networks. This new method is based on the GBC measure that is used to estimate the influence of a group of nodes over the information flow in the network. We show that the new method is $\left(1 - \frac{1}{e}\right)$ approximation for the problem of finding a set of *k* nodes with maximal contribution to the *GBC* of a given set of nodes *D*. The new method can be used to increase the deployment of a distributed monitoring system such as Distributed Network Intrusion Detection System on an evolving network.